\documentclass{nature}

\title{Interaction-driven quantum phase transition of a single magnetic impurity in Fe(Se,Te)}

\author{M. Uldemolins$^{1}$, A. Mesaros$^{1}$, G. D. Gu$^{2}$, A. Palacio-Morales$^{1}$, M. Aprili$^{1}$, P. Simon$^{1}$ \& F. Massee$^{1}$}

\usepackage{lineno}
\usepackage[english]{babel}
\usepackage{amssymb,amsmath}
\usepackage{graphicx}
\usepackage{multirow}
\usepackage{bigstrut}
\begin{document}
	
	\maketitle
	
	\begin{affiliations}
		\item Universit\'{e} Paris-Saclay, CNRS, Laboratoire de Physique des Solides, 91405, Orsay, France
		\item Condensed Matter Physics and Materials Science
		Department, Brookhaven National Laboratory, Upton, NY
		11973, USA\\
	\end{affiliations}
	
	\begin{abstract}
		Understanding the interplay between individual magnetic impurities and superconductivity is crucial for bottom-up construction of novel phases of matter. For decades, the description by Yu, Shiba and Rusinov (YSR) of single spins in a superconductor and its extension to include quantum effects has proven highly successful: the pair-breaking potential of the spin generates sub-gap electron- and hole excitations that are energetically equidistant from zero. By tuning the energy of the sub-gap states through zero, the impurity screening by the superconductor makes the ground state gain or lose an electron, signalling a parity breaking quantum phase transition. Here we show that in multi-orbital impurities, correlations between the in-gap states can conversely lead to a quantum phase transition where more  than one electron simultaneously leave the impurity without significant effect of the screening by the superconductor, while the parity may remain unchanged. This finding implies that the YSR treatment is not always valid, and that intra-atomic interactions, particularly Hund’s coupling that favours high spin configurations, are an essential ingredient for understanding the sub-gap states. The interaction-driven quantum phase transition should be taken into account for impurity-based band-structure engineering, and may provide a fruitful basis in the search for novel physics.
	\end{abstract}

To describe states appearing inside the energy gap of the superconductor at magnetic impurities \cite{heinrich_prog_2018, ji_prl_2008, yazdani_science_1997, franke_science_2011, menard_nphys_2015, hatter_ncom_2015, choi_ncomm_2017, huang_nphys_2020}, the standard Yu-Shiba-Rusinov (YSR) model \cite{yu, shiba, rusinov} represents the impurity as a single classical spin, whose pair-breaking potential leads to well defined electron- and hole excitations, called YSR states. A fundamental issue of the classical YSR impurity model, though, is that it cannot describe the true quantum ground state of the system \cite{matsuura, zitko}. This is problematic because a typical transition metal atom impurity adsorbed on a substrate is expected to exhibit quantum behavior, being far from the limit \cite{zitko} in which an effective YSR model could arise from an underlying quantum model. Even so, scanning tunneling spectroscopy (STS) experiments on various substrates and impurities\cite{yazdani_science_1997, ji_prl_2008, heinrich_prog_2018, franke_science_2011, menard_nphys_2015, hatter_ncom_2015, choi_ncomm_2017, huang_nphys_2020} have been successfully interpreted using YSR states. Only recently STS studies on superconductors have been used to extract parameters of an underlying, more complete quantum impurity model, namely, of the Kondo model \cite{kondo, matsuura, zitko}, which introduces a quantum impurity spin \cite{hatter_ncom_2015}, and the single-orbital Anderson Impurity Model (AIM), which adds quantum charge fluctuations on the impurity \cite{AstAIM}. Nevertheless, these works did not invalidate the applicability of the YSR paradigm. The great promise for bottom-up construction of novel phases of matter \cite{khajetoorians_review, kim_sciadv_2018, beck_naturecomm_2021} through manipulation of such impurities therefore urges a more fundamental understanding of the in-gap excitations and the associated quantum ground states.

Starting with the case of a single in-gap excitation due to a single impurity orbital, the quantum ground state of the system changes when the excitation energy crosses zero. This phenomenology is interpreted using a varying YSR exchange coupling between the classical impurity spin and the superconductor \cite{balatsky_revmodphys_2006}, even though YSR fails to describe any key feature of the two competing quantum ground states. Experimentally, quantum phase transitions (QPTs) have been chiefly observed for magnetic molecules whose exchange coupling with the substrate varies with adsorption site \cite{franke_science_2011, hatter_ncom_2015} and can even be tuned with the tip of a scanning tunnelling microscope \cite{farinacci_prl_2018}. None of these observations of a zero energy crossing, however, strictly require a quantum impurity model description. Interestingly, not only single, but also multiple in-gap excitation peaks are observed in various systems \cite{ji_prl_2008, franke_science_2011, choi_ncomm_2017}, implying that multiple impurity orbitals are likely involved, filled with electrons in a high-spin configuration. For transition metal impurities this is a natural consequence of a strong intra-atomic exchange, or Hund's coupling, which typically overcomes the crystal field splitting, hence the natural model is at least a multi-orbital AIM. Yet, these systems were successfully interpreted by adding up several independent classical YSR channels \cite{von_oppen_yu-shiba-rusinov_2021}. The question arises if one can observe a phenomenology that invalidates the interpretation through independent classical YSR channels and hence reveals the quantum interacting nature of the QPT of magnetic impurities on superconductors.

	\begin{figure}
		\centering
		\includegraphics[width=\textwidth]{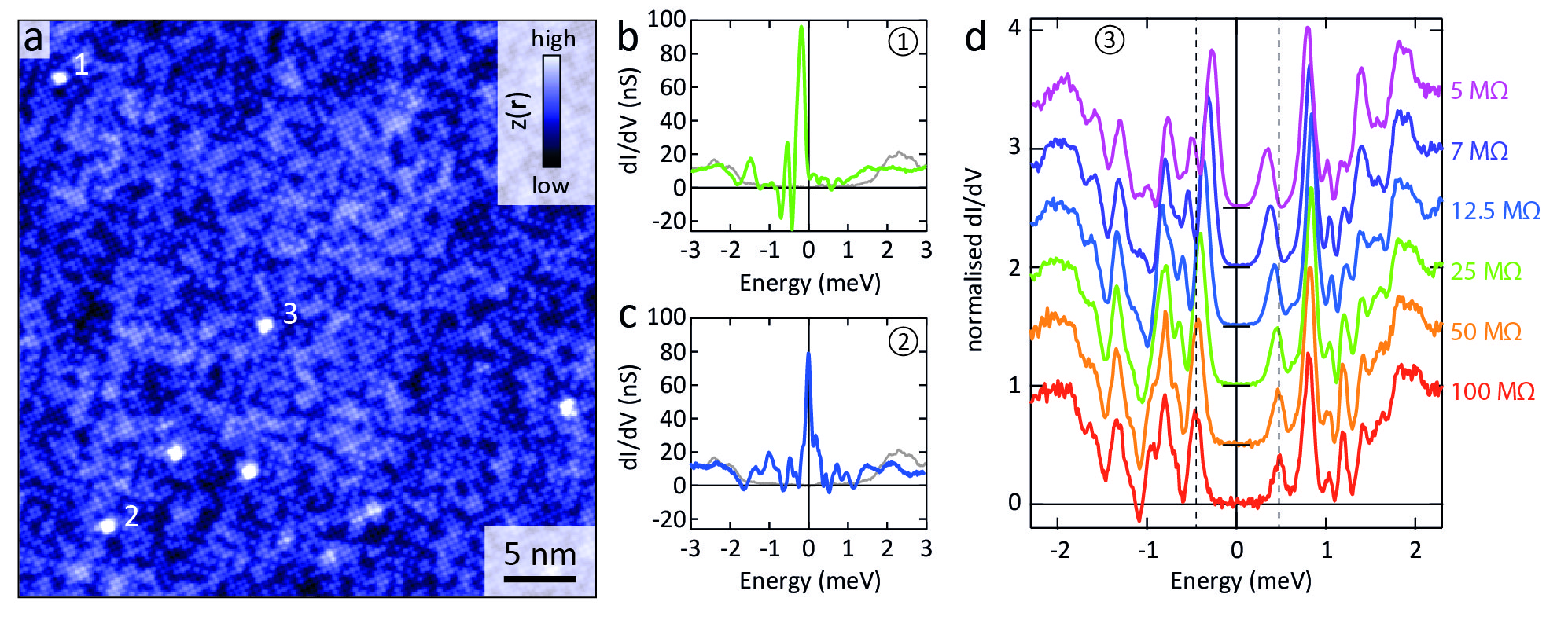}
  \caption{\label{fig:1} \textbf{Tunable multi-orbital YSR in Fe(Se,Te)}. \textbf{a} Constant current image of Fe(Se,Te) with several interstitial excess Fe atoms (bright protrusions). $V_{set} =$ 5~mV, $I_{set} =$ 100~pA. \textbf{b}, \textbf{c} Differential conductance spectra taken at the excess Fe impurities marked 1 and 2 in panel a, respectively (setup resistance = 100~M$\Omega$). Both show a multitude of peaks highlighting the multi-orbital nature of the Fe sites. The appearance of negative differential conductance in both spectra reflects interactions between sub-gap states. The fully gapped grey spectrum is taken in between impurities 1 and 3 for comparison. \textbf{d} Height dependence of differential conductance on the Fe impurity marked 3 in panel a. Upon lowering the setup resistance, the peaks shift closer to zero.}
	\end{figure}

To search for tunable, interacting in-gap states, we studied excess Fe atoms at the surface of superconducting FeSe$_{0.45}$Te$_{0.55}$, whose in-gap states were recently shown to be sensitive to the tip-sample distance \cite{leiden_2021}. Fe(Se,Te) thus combines the ability to tune the sub-gap excitation energies, with the high-spin configuration of an interstitial Fe impurity resulting from Hund's coupling of electrons in the d-orbitals. Figure \ref{fig:1}a shows a typical constant current image of cleaved Fe$_{1+x}$Se$_{0.45}$Te$_{0.55}$ containing several excess Fe impurities recognisable as bright protrusions among the mixture of respectively bright and dark Te and Se atoms. The excess Fe concentration (x $\sim 0.05\%$) is chosen to be high enough to have ample choice of local interstitial Fe environment, yet still low enough to observe a fully gapped superconducting spectrum in between the impurities. Contrary to previous reports \cite{yin_nphys_2015, wang_science_2020}, all of the up to 100 excess Fe we probed show multiple in-gap peaks. The energy of the in-gap peaks varies for different impurities, highlighting the variation in coupling parameters with local environment of e.g. Se and Te mixture. Differential conductance spectra taken at the centre of two different Fe impurities are shown in Fig.~\ref{fig:1}b, c. Interestingly, not only peaks are observed, but a large fraction of excess Fe also displays negative differential conductance (NDC) at one or more energies. Whereas NDC is common for superconducting tips\cite{franke_science_2011, hatter_ncom_2015, choi_ncomm_2017, huang_nphys_2020, heinrich_prog_2018, ji_prl_2008, farinacci_prl_2018, leiden_2021}, for the normal metal tip we use here the only possibility for NDC to occur within the fully gapped superconductor is through interactions between the different sub-gap states \cite{thielmann_super-poissonian_2005}. Last but not least, by changing the tip-sample distance all excitation energies can be shifted as shown in Fig.~\ref{fig:1}d.
	
	\begin{figure}
		\centering
		\includegraphics[width=0.7\textwidth]{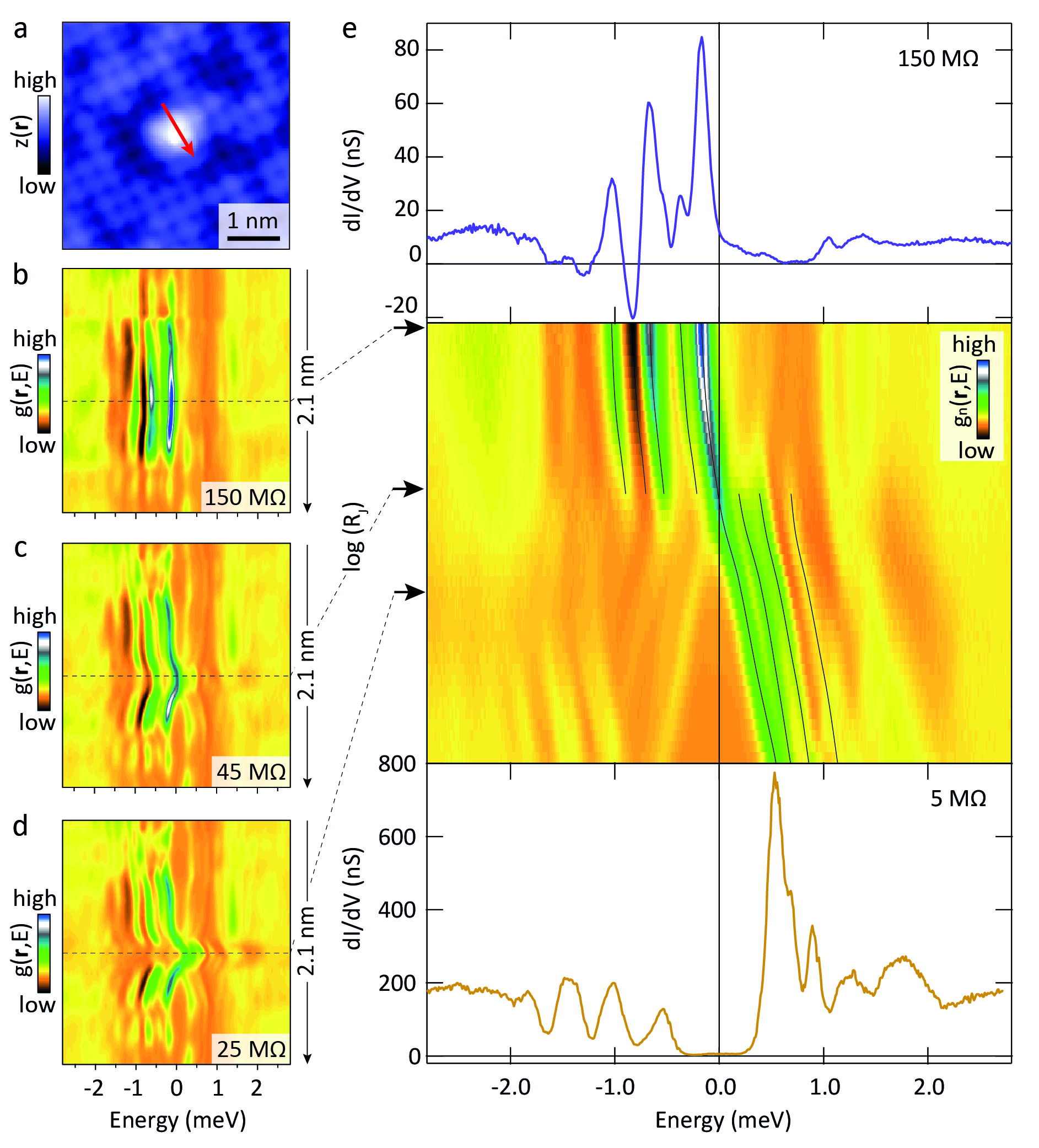}
  
  \caption{\label{fig:2} \textbf{Crossing zero energy}. \textbf{a} Constant current image of FeSe$_{0.45}$Te$_{0.55}$ with an excess Fe atom in the middle. \textbf{b}-\textbf{d} Differential conductance taken along the line in panel a for 150~M$\Omega$, 45~M$\Omega$ and 25~M$\Omega$, respectively. Directly on top of the Fe impurity the in-gap states shift most dramatically and cross zero. \textbf{e} Junction resistance dependence of the normalised differential conductance directly above the Fe impurity ranging from 150~M$\Omega$ (top) to 5~M$\Omega$ (bottom). The step size in resistance uses a logarithmic scale, the black lines are guides to the eye. As soon as the first state crosses zero, the intensity of all others is affected. We note that since the spectra are normalised outside the gap and since there is no dramatic change in intensity without a crossing (Fig.~\ref{fig:1}d) or of the state crossing zero, the intensity switch of the other states is not an artifact of normalization. Disappearance of NDC can moreover not be due to normalisation, and un-normalized spectra in panel d also clearly show the switch.}
	\end{figure}
	
Having identified controllable, multiple in-gap states, we can next use the tip-impurity distance to shift low-lying states through zero and track their evolution. To determine the lateral range of the tip-induced shifts, and to find the location with the most dramatic shift, we first take a line cut across the excess Fe atom in Fig.~\ref{fig:2}a at different junction resistances. As Fig.~\ref{fig:2}b shows, at high junction resistance (large tip-sample distance), the in-gap states hardly shift as function of position. Upon nearing the tip to the surface, however, the states move towards zero energy (Fig.~\ref{fig:2}c), and the first one eventually crosses when the tip is directly above the Fe impurity and at a sufficiently low junction resistance (Fig.~\ref{fig:2}d). To analyse the crossing more precisely, we perform a detailed junction resistance dependence on top of the Fe impurity, see Fig.~\ref{fig:2}e, using resistances ranging from 150~M$\Omega$ (Fig.~\ref{fig:2}e, top) to 5~M$\Omega$ (Fig.~\ref{fig:2}e, bottom). Since the electric field is linearly proportional to the tip-sample distance, and the current (and thus resistance) depends exponentially on the distance, we furthermore use a logarithmic scale for the resistance to enhance visibility of the crossing. For large tip-sample distance, the in-gap state closest to zero energy is strongest in intensity at negative bias, and is followed by additional peaks and a relatively prominent negative differential conductance dip (similar to the Fe in Fig.~\ref{fig:1}b). Importantly, the intensity of all peaks is stronger at negative bias than at positive bias. For lower setup resistances, all features shift closer to zero bias. Then, as soon as the first peak crosses zero bias, all other peaks switch polarity (their intensity becomes dominant at positive bias instead of negative bias), while the NDC disappears. Additionally, whereas the peak that crossed continues to shift to higher energy, the others never cross, but instead an ever increasing gap is formed after the crossing point. For independent in-gap states this would not have been possible: higher-energy states would simply have continued shifting until they also would have crossed zero. 

To understand the observed behaviour we focus on explaining two key observations: (1) the simultaneous switch of the in-gap states from hole-like (dominant intensity at $E<0$) to electron-like (dominant intensity at energy $E>0$), and (2) the appearance of negative differential conductance. The latter phenomenon occurs even though the experiment is well within the tunneling regime, implying that the tunneling current cannot be described by the single-particle density of states: the impurity must involve at least two states with variable electron occupation, which allows the blocking of tunneling current (and hence negative conductance) through one state due to interactions with electrons present in the other state \cite{thielmann_super-poissonian_2005}. Therefore, the impurity cannot be modeled by independent (classical or quantum) spins.

 \begin{figure}
  \centering
\includegraphics[width=\textwidth]{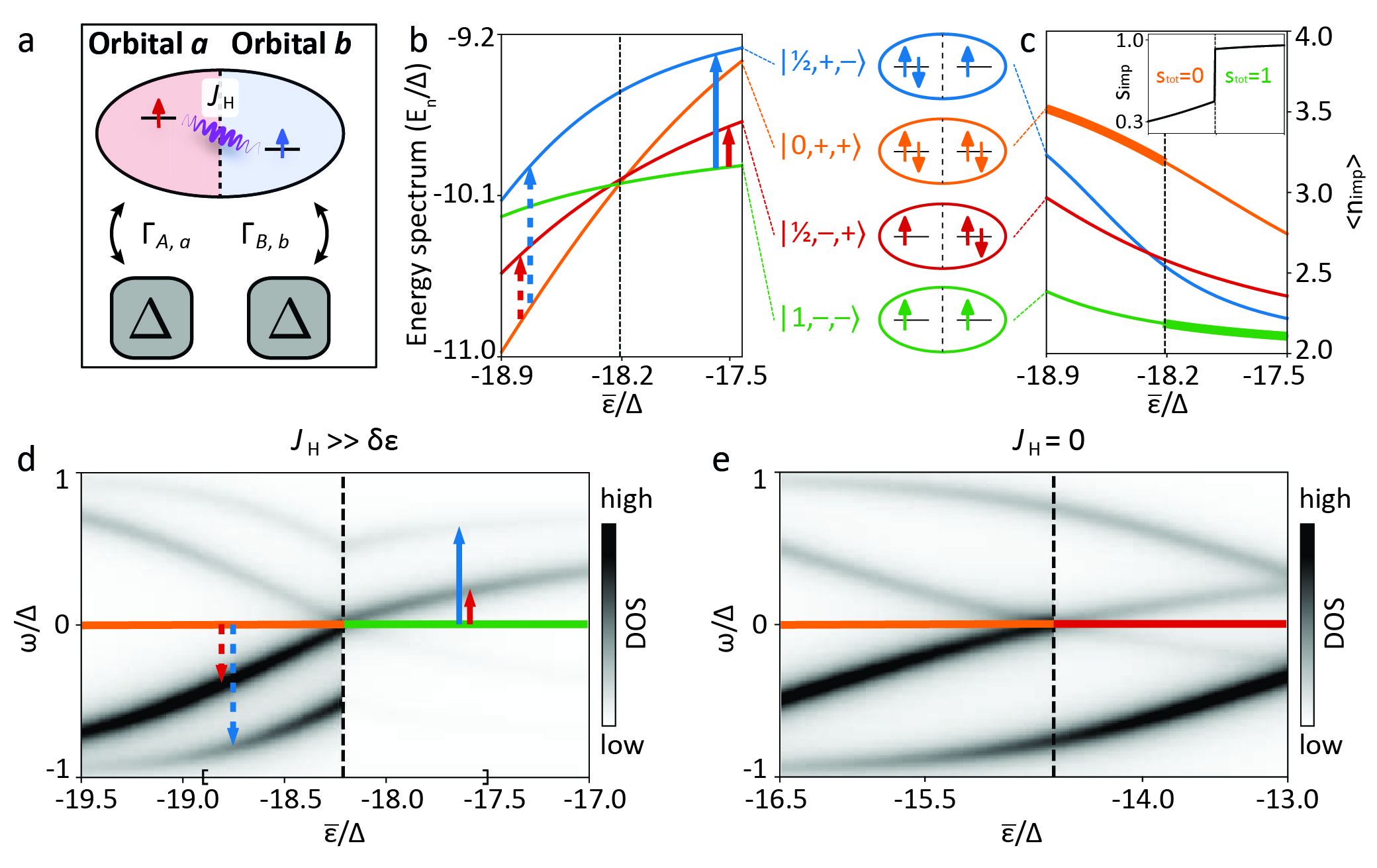}
\caption{\label{fig:theo} \textbf{Theoretical description of a multi-channel QPT}. \textbf{a} Sketch of the model Hamiltonian. The ellipse represents an impurity with two orbitals interacting via a Hund's coupling $J_{\mathrm{H}}$, while squares represent the two scattering channels of the superconductor coupled to the orbitals, each channel modeled by a superconducting site. \textbf{b} Evolution of the energy of the four lowest-lying many-body states for increasing average orbital energy $\overline{\varepsilon}/\Delta$ driven by the STM tip electric field. The vertical arrows indicate the STM-accessible excitations before and after the MCQPT. All the energy curves were shifted by $-2\overline{\varepsilon}/\Delta$ to improve readability. \textbf{c} Evolution of the average total electron occupation on the impurity for all four many-body states (same coloring as in panel b). The thick line marks the ground state for the given value of $\overline{\varepsilon}/\Delta$. The inset shows the total spin $s_{\mathrm{tot}}$ of the ground state in the same $\overline{\varepsilon}/\Delta$ range. \textbf{d} Simulated in-gap LDOS at the impurity site as a function of $\overline{\varepsilon}/\Delta$. Vertical arrows same as panel b, where now the direction marks the dominant electron- or hole-nature. The region of $\overline{\varepsilon}/\Delta$ shown in b and c is delimited by the square brackets on the horizontal axis. \textbf{e} Simulated in-gap LDOS for the transition in case of absence of Hund's coupling, $J_{\mathrm{H}}=0$. The color-coding of the horizontal line at $\omega = 0$ in panels d, e represents the ground state (colors from panels b, c). The vertical dashed lines in panels b-e mark the quantum phase transition. Calculation parameters in b, c, d: $J_{\mathrm{H}}=30\Delta$, $U = 15\Delta$, $\Gamma_{A,a} = \Gamma_{B,b} = 5\Delta^2$, $\delta \varepsilon = -6\Delta$. Same for panel e except $J_{\mathrm{H}}=0$ and $\delta \varepsilon = -3\Delta$. Phenomenological broadening in d, e is $\eta = 0.075\Delta$.}
\end{figure}

As a minimal model (Fig.~\ref{fig:theo}a) we consider a two-orbital Anderson impurity coupled to a superconducting bath. As we concentrate exclusively on in-gap states, we simplify the problem by treating the substrate as an $s$-wave superconductor with zero-bandwidth, where the BCS density of states is replaced by two quasi-particle states at energy $\Delta$. This approximation is known to perform very well for in-gap bound states of spin impurities \cite{von_oppen_yu-shiba-rusinov_2021}. To address the necessary interactions, we take each orbital as characterized by the impurity energy level $\varepsilon_{\alpha= a, b}$, and intra-orbital Coulomb energy $U_{\alpha= a,b} \equiv U$, and we assume they are each coupled to a different site of the superconducting bath $i = A, B$ via tunneling rates $\Gamma_{A,a}$, $\Gamma_{B,b} > \Delta^2$. The combination of site and orbital $(i,\alpha)=(A,a)\textrm{ or }(B,b)$ is referred to as a channel, since the $i$ state is a scattering channel in the superconductor that best couples with the orbital $\alpha$. The key interaction that couples the orbitals can be represented as the Hund's coupling,
\begin{equation}
	H_{J} =  - J_{\mathrm{H}} \; \mathbf{S}_a\cdot \mathbf{S}_b,
\end{equation}
with $\mathbf{S}_{\alpha=a,b}$ the spin operator for each orbital (see Fig.~\ref{fig:theo}a and Methods). To favour higher-spin configurations typical of transition-element impurities, we set the phenomenological parameters as
\begin{equation}
  \label{eq:theo_params}
J_{\mathrm{H}} > |\delta \varepsilon| \equiv |\varepsilon_{b}-\varepsilon_{a}| > \Delta.  
\end{equation}
Previous work on sub-surface impurities in FeTe\textsubscript{0.55}Se\textsubscript{0.45} indicate that the impurity states can be affected by the electric field of the tip \cite{leiden_2021}, likely due to the partial screening associated to a low super-fluid density \cite{homes_fete_2015}. The excess Fe atoms we study are on top of the Se/Te surface layer and therefore expected to be even more susceptible since they are closer to the tip. We therefore assume that any movement of the tip affects exclusively the energy levels of the impurity orbitals. We stress that this is a distinctly different mechanism for shifting in-gap states than is the case for molecules \cite{franke_science_2011, hatter_ncom_2015, farinacci_prl_2018}, where the presence of the tip modifies the hybridization of the impurity with the surrounding substrate. Since the electric field, which depends linearly on the distance, is driving the shift in energy levels, we assume that the average impurity energy level $\overline{\varepsilon}$ linearly varies with the tip-sample distance in accord with a previous treatment of impurity-bound states in this compound \cite{leiden_2021}. The $\overline{\varepsilon}$ thereby controls the energy of the sub-gap excitations and represents the key experimentally tunable parameter. We further assume for simplicity that the difference between energy levels $\delta \varepsilon$ remains constant.

In this model, we discover a multi-channel quantum phase transition (MCQPT) in the evolution of the four lowest-lying many-body eigenstates. We can label the eigenstates  as $|s_{\mathrm{tot}}, p_{A,a}, p_{B,b}\rangle$, due to the conserved total-spin $s_{\mathrm{tot}}$, and conserved electron-number parity $p_{i,\alpha}$ of channels $(i,\alpha)$ (see Methods). The four states are $|0, +, +\rangle$, $|1/2, +, -\rangle$, $|1/2, -, +\rangle$, $|1, -, -\rangle$, with $\pm$ denoting even and odd parity, respectively. The evolution of the energies of the four states with $\overline{\varepsilon}$ is shown in Fig.~\ref{fig:theo}b. The phase transition occurs as the state $|0, +, +\rangle$ exchanges places with the state $|1, -, -\rangle$. Hence at the transition both channels are involved, and both undergo a parity flip, so that the total parity does not change. This property is in stark contrast with the generic spin-impurity model quantum phase transition (e.g. associated to the Shiba model), but is actually not the universal signature of our MCQPT. Namely, we find that by adding a third orbital to our model, the same complete phenomenology can be recovered even though the parity flip of all three channels makes the total parity change sign across the transition. The essential property of our MCQPT is the large change of the average orbital occupation, $\langle n_{\mathrm{imp}}\rangle$, from nearly 4 (both orbitals fully occupied) to nearly 2 (both singly occupied) as $\overline{\varepsilon}$ changes, see Fig. ~\ref{fig:theo}c. The distinctive switch of all in-gap excitations from hole-like to electron-like is due to the depletion of the average electron occupation of the impurity orbitals. It is the strong Hund's coupling that determines the MCQPT to be from fully occupied orbitals and a low-spin state (total spin of system $s_{\mathrm{tot}} = 0$) into singly occupied orbitals and a high-spin state (total spin of system $s_{\mathrm{tot}} = 1$). This mechanism is allowed by a reasonable assumption that the impurity is in a mixed-valence state, i.e., the energy cost of the change of orbital occupation can be compensated by the gain of Hund's energy, leading to the requirement
\begin{equation}
  \label{eq:theo_param_JH}
J_H\geq U\sim|\overline{\varepsilon}|.  
\end{equation}
To address the STM observations, we calculate the total local density of states (LDOS) of the impurity orbitals (see Methods). Since we consider single-particle excitations, only transitions from the many-body ground state to states with opposite \textit{total parity} can have a non-zero spectral weight. To demonstrate the phenomenology of the MCQPT we focus on the two lowest in-gap excitations which result from the four lowest many-body eigenstates described above (see arrows in Fig.~\ref{fig:theo}b, d). To be precise, before the MCQPT, the in-gap excitations correspond to excitations from $|0, +, +\rangle$ to $|1/2, -,+\rangle$ and to $|1/2, +,-\rangle$; there is a $+ \rightarrow -$ parity flip in channels $(A,a)$ or $(B,b)$, respectively. After the MCQPT, they correspond to excitations from $|1, -, -\rangle$ to $|1/2, -,+\rangle$ and  $|1/2, +,-\rangle$. Panel Fig.~\ref{fig:theo}d shows clear agreement with the STM results (Fig.~\ref{fig:2}e), where the MCQPT is signaled by the lowest excitation crossing zero energy, and there is a concurrent change of spectral weight in the higher excitation.

Another key feature of our model is that the single-particle excitations appearing in the in-gap LDOS result from different eigenstates before and after the MCQPT. Therefore, in general there is a discontinuity of slope of the in-gap state energy at the point it crosses zero-bias with changing $\overline{\varepsilon}$ (see Fig.~\ref{fig:theo}d and Fig.~S2a,c in Supplementary Note 1). Indeed, this property is a common feature for the experimentally measured Fe impurities where a MCQPT occurs (see Supplementary Note 1), further supporting our interpretation. Note that the discontinuity of the slope at the MCQPT is a priori independent of the precise functional dependence of $\overline{\varepsilon}$ on the electric field.

Importantly, the phenomenology discussed above is not fine-tuned, but is present for a range of values of parameters as long as we stay in the regime given by Eqs.~\eqref{eq:theo_params},\eqref{eq:theo_param_JH}, as we show through a systematic study in Supplementary Note 1. In Supplementary Note 2 we also relate the excitation spectrum to a transport calculation to demonstrate that, without changing any of the above conclusions, negative differential conductance can occur for a reasonable range of asymmetry between the tip and the wave function of the bound states as is known in the case of tunneling through a quantum dot with multiple interacting levels \cite{thielmann_super-poissonian_2005}. To confirm our interpretation and the significance of multi-orbital interactions, we present the calculated LDOS with the Hund's coupling switched off ($J_{\mathrm{H}}=0$) in panel Fig.~\ref{fig:theo}e, where obviously the experimental phenomenology is not captured. We note that the $J_{\mathrm{H}}=0$ model completely decouples the two orbitals, hence in accord with intuition the spectral weight of the excitation in one channel is not affected by the transition in the other channel (see Supplementary Note 3). The same negative outcome is obtained for independent Kondo channels, or independent YSR channels. 

Despite the inability to reproduce negative differential conductance, could the switching of intensity be recovered with interlocked excitations in YSR or Kondo models? Multiple in-gap excitations may appear through various mechanisms in impurity models, e.g. due to a higher impurity spin (classical \cite{ruby_orbital_2016, moca_spin-resolved_2008, arrachea_yu-shiba-rusinov_2021} or quantum \cite{von_oppen_yu-shiba-rusinov_2021}), splitting by anisotropy \cite{zitko, hatter_ncom_2015}, and in e.g. FeSe due to multiple superconducting gaps \cite{li_magnetic_2009}. Notably among these, in a model of a quantum higher spin on the impurity, there may be total-parity-conserving QPT in which two channels undergo a screening transition simultaneously, either due to anisotropy, or due to symmetry-protected degeneracy of some impurity-substrate couplings. In such a transition, the spectral weights of some in-gap states may shift between hole- and electron-like, although it is not evident that such a shift can occur for excitations that do not cross zero. Furthermore, this is a rather fine-tuned situation not expected to hold for our myriad of impurities.

We note that a realistic description of a Fe impurity demands a larger number of orbitals, which in turn, would yield a larger number of in-gap states; however, the two-orbital model discussed here contains the minimal ingredients to capture the distinctive feature of the MCQPT. We further note that in our model the transition is not at all characterized by a change in screening of impurity by substrate, instead it is characterized by a change of occupation and total spin driven by Hund's interaction. Lastly, our model with its single-site single-gap $s$-wave superconductor obviously cannot address the spatial features of impurity-bound states observed by horizontal movement of the tip. However, the fact that the key features in experiment are found at various distances from the impurity in both horizontal and vertical movement indicates that we are observing universal features driven by the field effect, and not fine-tuned features due to spatial- or momentum-dependent properties of the impurity and substrate.

The experimental observation of interacting sub-gap states in superconducting Fe(Se,Te) challenges the validity of the celebrated YSR model in this system. Given the prevalence of using multi-orbital transition-element atoms for bottom-up construction of devices \cite{khajetoorians_review, heinrich_prog_2018, kim_sciadv_2018, beck_naturecomm_2021}, this may furthermore be true for a wide range of other systems. Where the YSR model fails, our alternative, multi-orbital model that takes intra-orbital interactions into account accurately describes the experimentally observed characteristics. Whereas the additional interactions may complicate the theoretical description of individual and coupled magnetic impurities, the extra parameter could perhaps be experimentally exploited to generate previously unanticipated phenomena as exemplified by the multi-orbital quantum phase transition.
	
	\begin{methods}
		\textbf{Tip and sample preparation.} Fe(Se,Te) single crystals were grown using the self-flux method. As-grown samples with a superconducting transition temperature of 14.5~K were used throughout this work. The crystals were mechanically cleaved in cryogenic vacuum at T $\sim$ 20~K and directly inserted into the STM head at 4.2~K. An etched atomically sharp and stable tungsten tip was used for all measurements. Differential conductance measurements were performed by numerical derivation as well as with a standard lock-in amplifier operating at 429.7Hz. All measurements were recorded at the base temperature of T = 0.3~K.
		
		\textbf{Theory.} The total Hamiltonian reads
\begin{equation}
	\label{eq:ham}
	H  = H_{\mathrm{SC}} + H_{\mathrm{imp}} + H_{\mathrm{T}},
\end{equation}
with
\begin{align}
	H_{\mathrm{SC}} &= \sum_i\left(\Delta c^{\dagger}_{i,\uparrow}c^{\dagger}_{i,\downarrow} + \mathrm{h.c.}\right),\\
	H_{\mathrm{imp}} &= \sum_{\alpha, \sigma} \varepsilon_{\alpha} \hat{n}_{\alpha \sigma} + \sum_{\alpha} U_{\alpha} \hat{n}_{\alpha,\uparrow}\hat{n}_{\alpha,\downarrow} - J_{\mathrm{H}} \; \mathbf{S_a}\cdot \mathbf{S_b},\\
	H_{\mathrm{T}} &= \sum_{i,\alpha,\sigma} \left(t_{i,\alpha} c^{\dagger}_{i,\sigma}d_{\alpha \sigma} + \mathrm{h.c.}\right),
\end{align}
where the operators $d^{\dagger}_{\alpha, \sigma}$ create a spin-$\sigma$ electron in impurity orbital $\alpha=a,\; b$, each coupled to a superconducting site $i=A, \;B$, respectively, on which the electron creation operator is $c^{\dagger}_{i, \sigma}$. The $\hat{n}_{\alpha,\sigma} = d^{\dagger}_{\alpha, \sigma}d_{\alpha, \sigma}$ represents the particle-number operator on the impurity (not a conserved quantity), while the impurity spin operator reads
\begin{equation}
	\mathbf{S}_{\alpha} = \sum_{\sigma, \sigma'} d^{\dagger}_{\alpha, \sigma}\left(\sigma_x, \sigma_y, \sigma_z\right)_{\sigma, \sigma'}d_{\alpha,\sigma'}.
\end{equation}

The operators for \textit{total spin}, \textit{channel-1 parity} and \textit{channel-2 parity}, all conserved quantities introduced to label the eigenstates, read
\begin{align}
		\mathbf{S}_{\mathrm{tot}} &= \left(\sum_{i}\mathbf{S}_i+\sum_{\alpha}\mathbf{S}_{\alpha}\right),\\
		P_{A,a} &= (-1)^{\hat{n}_{A,a}},\\
		P_{B,b} &= (-1)^{\hat{n}_{B,b}},
\end{align}
where $\hat{n}_{i,\alpha} = \sum_{\sigma}(c^{\dagger}_{i, \sigma}c_{i, \sigma} + d^{\dagger}_{\alpha, \sigma}d_{\alpha, \sigma})$. Owing to the Hamiltonian's full spin-rotation symmetry, eigenstates can also be labeled according to the $z$ component of the total spin $m$, but we dropped this label to lighten the notation in the main text. Note that eigenstates $|0, +, +\rangle$, $|1/2, +, -\rangle$, $|1/2, -, +\rangle$, $|1, -, -\rangle$ are indeed zero-, two-, two- and three-fold degenerate, respectively. Further, the \textit{total Fermion parity} operator $P = P_{A,a} \cdot P_{B,b}$ also commutes with the total Hamiltonian.

The LDOS on the impurity site reads
\begin{equation}
	\rho_{\mathrm{imp}}(\omega) = -\frac{1}{\pi}\operatorname{Im}\sum_{\alpha, \sigma}G^{\mathrm{R}}_{\alpha, \sigma; \alpha, \sigma}(\omega),
\end{equation}
where the retarded Green function can be expressed through the Lehmann representation at $T=0$,
\begin{equation}
\label{eq:gfr_0T}
	G^{\mathrm{R}, T=0}_{\alpha, \sigma; \alpha, \sigma}(\omega) = \frac{1}{d_{\mathrm{GS}}}\sum_{\mathrm{GS}}\sum_{n} \left[\frac{|\langle \mathrm{GS}|d^{\dagger}_{\alpha,\sigma}|n\rangle|^2}{\omega+E_n-E_{\mathrm{GS}}+i0^{+}} + \frac{|\langle n|d^{\dagger}_{\alpha,\sigma}|\mathrm{GS}\rangle|^2}{\omega+E_{\mathrm{GS}}-E_{n}+i0^{+}},\right],
\end{equation}
with $d_{\mathrm{GS}}$ the degeneracy of the ground state. The experimental temperature is low, namely, the smallest energy scale in our problem, $k_{\mathrm{B}}T \ll \Delta$. In this case, the contributions to the LDOS from transitions between excited states are exponentially small, and they would only induce a slight broadening at the level crossings. Hence in the modeling we assume $T=0$. The tunneling rate introduced in the main text is related to the tunneling amplitude as $\Gamma_{i,\alpha} = \pi t_{i,\alpha}^2$.
	
	\end{methods}

	\begin{addendum}
		\item We thank G. Zar\'{a}nd, C. Pepin and Q. Si for useful discussions. M.U., A.M. and P.S. acknowledge the support of the French Agence Nationale de la Recherche (ANR), under grant number ANR-22-CE30-0037. The work at BNL was supported by the US Department of Energy, office of Basic Energy Sciences, contract no. DOE-sc0012704. FM would like to acknowledge funding from the ANR (ANR-21-CE30-0017-01).
		\item[Author contributions] F.M., A.M. and P.S. supervised the project. F.M. conceived and performed the experiment, and analysed the data. M.U. performed the theoretical study with A.M. and P.S.. Samples were synthesized by G.D.G.. The manuscript reflects contributions from all authors.
		\item[Competing Interests] The authors declare that they have no competing financial interests.
		\item[Correspondence] Correspondence and requests for materials should be addressed to F.M. \\
		(email: freek.massee@universite-paris-saclay.fr).
	\end{addendum}


\begin{thebibliography}{}

		\bibitem{yazdani_science_1997}
		A. Yazdani, B. A. Jones, C. P. Lutz, M. F. Crommie and D. M. Eigler, Science \textbf{275}, 1767 (1997) \textit{Probing the local effects of magnetic impurities on superconductivity}

        \bibitem{ji_prl_2008}
		S. -H. Ji, T. Zhang, Y. -S. Fu, X. Chen, X. -C. Ma, J. Li, W. -H. Duan, J. -F. Jia and Q. -K. Xue, Physical Review Letters \textbf{100}, 226801 (2008) \textit{High-resolution scanning tunneling spectroscopy of magnetic impurity induced bound states in the superconducting gap of Pb thin films}
		
		\bibitem{franke_science_2011}
		K. J. Franke, G. Schulze, and J. I. Pascual, Science \textbf{332}, 940-944(2011) \textit{Competition of superconducting phenomena and Kondo screening at the nanoscale}
		
		\bibitem{menard_nphys_2015}
		G. C. M\'enard, S. Guissart, Ch. Brun, S. Pons, V. S. Stolyarov,
		F. Debontridder, M. V. Leclerc, E. Janod, L. Cario, D. Roditchev,
		P. Simon and T. Cren, Nature Physics \textbf{11}, 1013 (2015) \textit{Coherent long-range magnetic bound states in a superconductor}
		
		\bibitem{hatter_ncom_2015}
		N. Hatter, B. W. Heinrich, M. Ruby, J. I. Pascual and K. J. Franke, Nature Communications \textbf{6}, 8988 (2015) \textit{Magnetic anisotropy in Shiba bound states across a quantum phase transition}
		
		\bibitem{choi_ncomm_2017}
		D. -J. Choi, C. Rubio-Verdu, J. de Bruijckere, M. M. Ugeda, N. Lorente and J. I. Pascual, Nature Communications \textbf{8}, 15175 (2017) \textit{Mapping the orbital structure of impurity bound states in a superconductor}
		
		\bibitem{huang_nphys_2020}
		H. Huang, C. Padurariu, J. Senkpiel, R. Drost, A. Levy Yeyati,
		J. C. Cuevas, B. Kubala, J. Ankerhold, K. Kern and C. R. Ast, Nature Physics \textbf{16}, 1227-1231 (2020) \textit{Tunnelling dynamics between superconducting bound states at the atomic limit}
		
		\bibitem{heinrich_prog_2018}
		B. W. Heinrich, J. I. Pascual and K. J. Franke, Progress in Surface Science \textbf{93}, 1-19 (2018) \textit{Single magnetic adsorbates on s-wave superconductors}
		
		\bibitem{yu}
		L. Yu, Acta Phys. Sin. \textbf{21}, 75 (1965) \textit{Bound state in superconductors with paramagnetic impurities}
		
		\bibitem{shiba}
		H. Shiba, Prog. Theor. Phys. \textbf{40}, 435 (1968) \textit{Classical spins in superconductors}
		
		\bibitem{rusinov}
		A. I. Rusinov, Pis’ma Zh. Eksp. Teor. Fiz. \textbf{9}, 146 (1968)
		[JETP Lett. \textbf{9}, 85 (1969)] \textit{Superconductivity near a paramagnetic impurity}
		
		\bibitem{matsuura}
		T. Matsuura, Prog. Theor. Phys. \textbf{57}, 1823–1835 (1977) \textit{The effects of impurities on superconductors with Kondo effect}
		
		\bibitem{zitko}
		R. Žitko, O. Bodensiek and T. Pruschke, Phys. Rev. B \textbf{83}, 054512 (2011) \textit{Effects of magnetic anisotropy on the subgap excitations induced by quantum impurities in a superconducting host}
		
		\bibitem{kondo}
		J. Kondo, Progr. Theor. Phys. \textbf{32}, 37-49 (1964) \textit{Resistance minimum in dilute magnetic alloys}

        \bibitem{AstAIM}
        H. Huang, R. Drost, J. Senkpiel, C. Padurariu, B. Kubala, A. Yeyati, J. Cuevas, J. Ankerhold, K. Kern and C. Ast, Communications Physics \textbf{3}, 199 (2020) \textit{Quantum phase transitions and the role of impurity-substrate hybridization in Yu-Shiba-Rusinov states.}

        \bibitem{khajetoorians_review}
		A. A. Khajetoorians, D. Wegner, A. F. Otte and I. Swart, Nature Review Physics \textbf{1}, 703–715 (2019) \textit{Creating designer quantum states of matter atom-by-atom}
	
		\bibitem{kim_sciadv_2018}
		H. Kim, A. Palacio-Morales, T. Posske, L. R\'{o}zsa, K. Palot\'{a}s, L. Szunyogh, M. Thorwart and R. Wiesendanger, Science Advances 4, eaar5251 (2018) \textit{Toward tailoring Majorana bound states in artificially constructed magnetic atom chains on elemental superconductors}
		
		\bibitem{beck_naturecomm_2021}
		P. Beck, L. Schneider, L. R\'ozsa, K. Palot\'as, A. L\'aszl\'offy, L. Szunyogh, J. Wiebe and R. Wiesendanger, Nature Communications \textbf{12}, 2040 (2021)
		\textit{Spin-orbit coupling induced splitting of Yu-Shiba-Rusinov states in antiferromagnetic dimers}

        \bibitem{balatsky_revmodphys_2006}
		A. V. Balatsky, I. Vekhter and J. -X. Zhu, Rev. Mod. Phys. \textbf{78}, 373 (2006) \textit{Impurity-induced states in conventional and unconventional superconductors}
  
		\bibitem{farinacci_prl_2018}
		L. Farinacci, G. Ahmadi, G. Reecht, M. Ruby, N. Bogdanoff, O Peters, B. W. Heinrich, F. von Oppen and K. J. Franke Phys. Rev. Lett. \textbf{121}, 196803 (2018) \textit{Tuning the coupling of an individual magnetic impurity to a superconductor: quantum phase transition and transport}

        \bibitem{von_oppen_yu-shiba-rusinov_2021}
        F. von Oppen and K. J. Franke, Phys. Rev. B, \textbf{103}, 205424 (2021) \textit{Yu-Shiba-Rusinov states in real metals}

        \bibitem{leiden_2021}
		D. Chatzopoulos, D. Cho, K. M. Bastiaans, G. O. Steffensen, D. Bouwmeester, A. Akbari, G. Gu, J. Paaske, B. M. Andersen and M. P. Allan, Nature Communications \textbf{12}, 298 (2021) \textit{Spatially dispersing Yu-Shiba-Rusinov states in the unconventional superconductor FeTe$_{0.55}$Se$_{0.45}$}
  
		\bibitem{yin_nphys_2015}
	    J. -X. Yin, Z. Wu, J. -H. Wang, Z. -Y. Ye, J. Gong, X. -Y. Hou, L. Shan, A. Li, X. -J. Liang, X. -X. Wu, J. Li, C. -S. Ting, Z. -Q. Wang, J. -P. Hu, P. -H. Hor, H. Ding and S. H. Pan, Nature Physics \textbf{11}, 543 (2015) \textit{Observation of a robust zero-energy bound state in iron-based superconductor Fe(Te,Se)}

		\bibitem{wang_science_2020}
		Z. Wang, J. Olivares Rodriguez, L. Jiao, S. Howard, M. Graham, G. D. Gu, T. L. Hughes, D. K. Morr and V. Madhavan, Science \textbf{367}, 104-108 (2020) \textit{Evidence for dispersing 1D Majorana channels in an iron-based superconductor}

        \bibitem{thielmann_super-poissonian_2005}
        A. Thielmann, M. H. Hettler, J. K\"onig, G. Sch\"on, Phys. Rev. B, \textbf{71}, 045341 (2005) \textit{Super-Poissonian noise, negative differential conductance, and relaxation effects in transport through molecules, quantum dots, and nanotubes}

        \bibitem{homes_fete_2015}
        C.C. Homes, Y. M. Dai, J. S. Wen, Z. J. Xu, G. D. Gu Phys. Rev. B \textbf{91} 144503 (2015) \textit{FeTe$_{0.55}$Se$_{0.45}$: A multiband superconductor in the clean and dirty limit}
 
        \bibitem{ruby_orbital_2016}
        M. Ruby, Y. Peng, F. von Oppen, B. W. Heinrich, K. J. Franke, Phys. Rev. Lett. \textbf{117} 186801 (2016) \textit{Orbital picture of Yu-Shiba-Rusinov Multiplets}

        \bibitem{arrachea_yu-shiba-rusinov_2021}
        L. Arrachea, Phys. Rev. B \textbf{104} 134515 (2021) \textit{Yu-Shiba-Rusinov multiplets and clusters of multiorbital adatoms in superconducting substrates: Subgap Green's function approach}

        \bibitem{moca_spin-resolved_2008}
        C. P. Moca, E. Demler, B. Jank\'o, G. Zar\'and Phys. Rev. B \textbf{77} 174516 (2008) \textit{Spin-resolved spectra of Shiba multiplets from Mn impurities in MgB$_2$}

        \bibitem{li_magnetic_2009}
        J. Li, Y. Wang EPL \textbf{88} 17009 (2009) \textit{Magnetic impurities in the two-band }$s_{\pm}$\textit{-wave superconductors}
		
	\end{thebibliography}
\end{document}

\maketitle
	
	\begin{affiliations}
		\item Universit\'{e} Paris-Saclay, CNRS, Laboratoire de Physique des Solides, 91405, Orsay, France
		\item Condensed Matter Physics and Materials Science
		Department, Brookhaven National Laboratory, Upton, NY
		11973, USA\\
	\end{affiliations}

\section*{Supplementary Note 1}
\begin{figure}
	\centering
	\includegraphics[width=\textwidth]{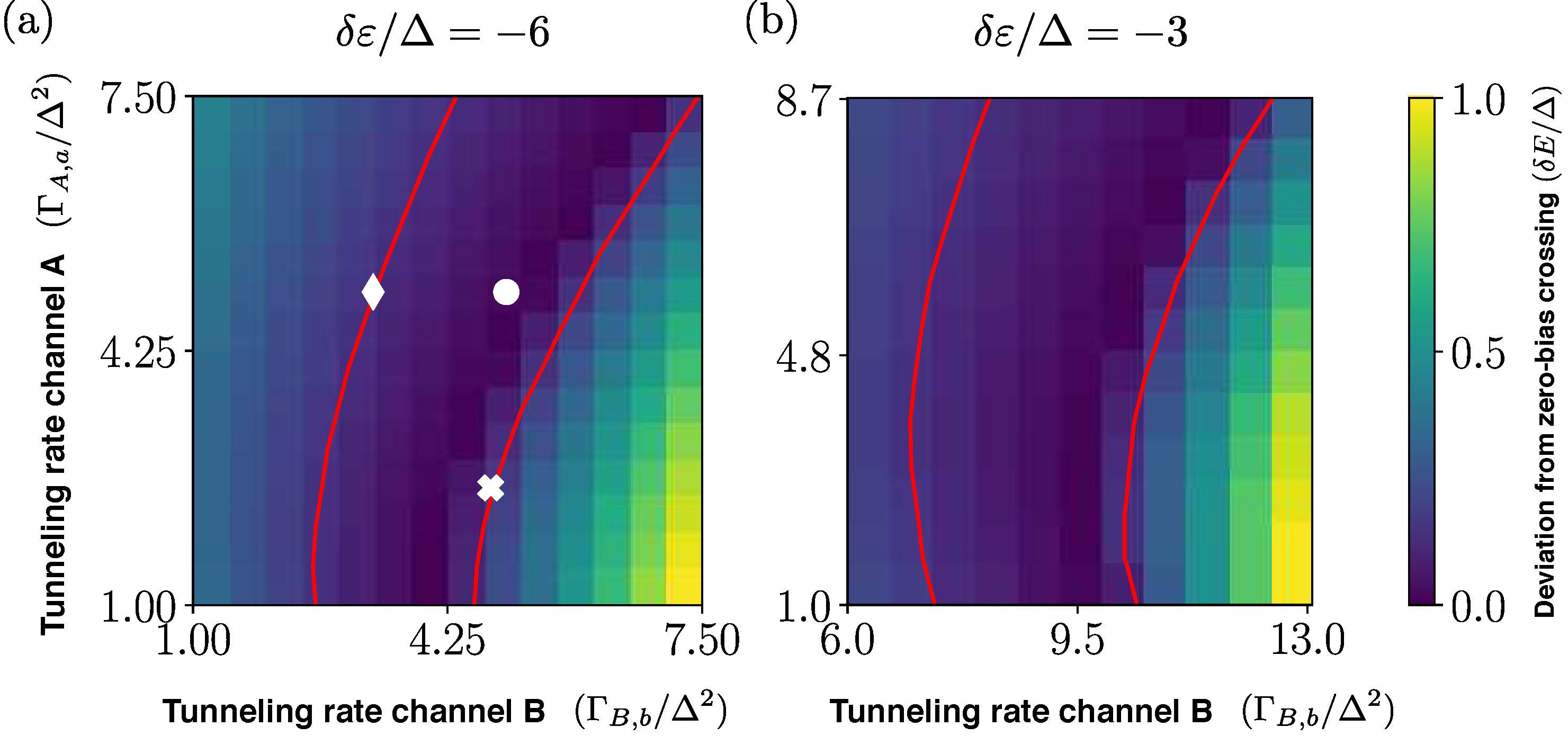}
	\caption{Magnitude of the gap in the lowest-lying excitation (i.e., deviation from a zero-bias crossing) at the MCQPT for \textbf{a}: $\delta \varepsilon/\Delta = -6$, and \textbf{b}: $\delta \varepsilon/\Delta = -3$. The rest of the parameters are the same in both plots ($U = 15$, $J=30$, in units of $\Delta$), but note that the axis span a slightly different range. The red contours correspond to $\delta E/\Delta = 0.13$, i.e. the experimental resolution for $\Delta = 1.5$ meV and electron temperature $T = 0.4$ K. The markers in (a) correspond to various choices of the tunneling rates presented in the text: Fig.~3 in the main text (circle) and Fig.~S\ref{fig:supp_spec} (diamond and cross).}
	\label{fig:supp_gamma}
\end{figure}

\begin{figure}
	\centering
	\includegraphics[width=\textwidth]{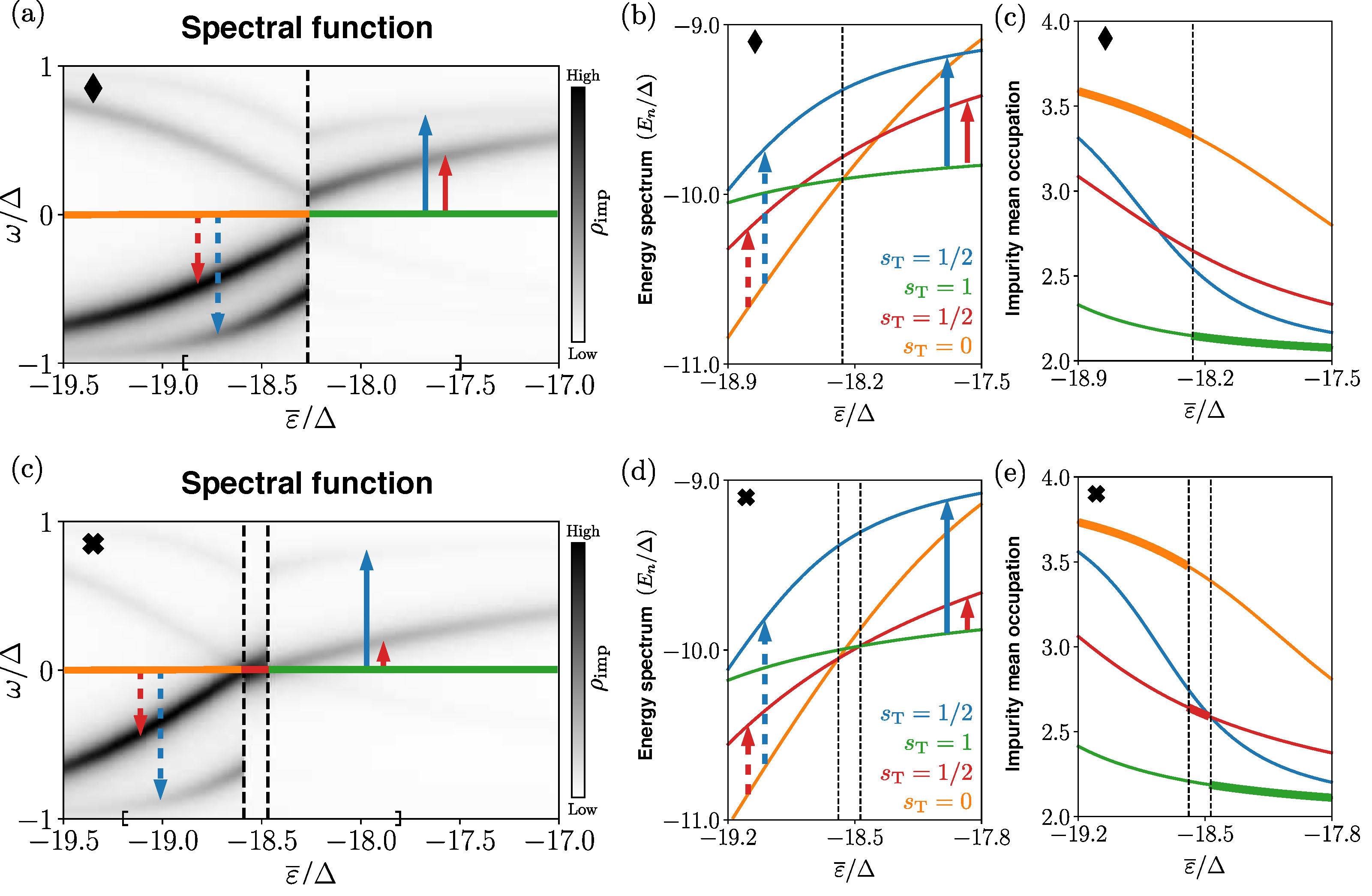}
	\caption{\textbf{Robustness of the MCQPT}. Top row: \textbf{a} Simulated in-gap LDOS at the impurity site as a function of $\overline{\varepsilon}/\Delta$. \textbf{b} Evolution of the energy of the four lowest-lying many-body states. \textbf{c} Evolution of the average total electron occupation on the impurity for all four many-body state. The color-coding and the arrows are the same as in main text  Fig.~3. The parameters correspond to the diamond marker in Fig.~S\ref{fig:supp_gamma}. Bottom row: Same as top row but with tunneling rates corresponding to the cross marker in Fig.~S\ref{fig:supp_gamma}. In this case, the deviation from zero-bias crossing $\delta E/\Delta$ is defined as the separation of the two QPT (black dashed lines), which yields a discontinuity in the in-gap excitations.}
	\label{fig:supp_spec}
\end{figure}

In this Supplementary Note we discuss the robustness of the observations made in the main text to variations in the model parameters and show additional experimental observations. We start by noting one consequence of the nature of the new multi-channel QPT: since the excitations involve transitions between different states before and after the QPT (and not a simple crossing as in the classical Shiba model), the lowest-lying excitation does not need to cross zero-bias energy at the transition. Hence, the crossing in the theoretical model is not protected but depends on the parameters. Experimentally, the several examples of transitions we identify all show the crossing, at least within the experimental energy resolution. Hence in particular for main text Fig.~3 of the main text we choose to present a situation where the energy of the first excitation satisfies that $E_{|1/2,-,+\rangle} - E_{\mathrm{GS}} \simeq  0$ at the MCQPT, and there is a zero-bias crossing in the in-gap LDOS. Nevertheless, we show in Fig.~S\ref{fig:supp_gamma} that there exists a wide region in the parameter space, within the already introduced model assumptions of main text Eqs.~(2),(3), for which the gap in the lowest-lying-excitation energy is below the experimental resolution. As the figure shows, the requirement of observing the crossing within resolution imposes a quite natural condition that the tunneling rates for the two orbitals are of the same order of magnitude, and the coupling of orbital to substrate is larger than the superconducting gap. In Fig.~S\ref{fig:supp_spec} (a-c) we give an example of the opposite scenario, where the MCQPT clearly lacks a zero-energy crossing. We emphasize that our MCQPT in absence of crossing still has the other essential features: the large change of the average orbital occupation at the transition, which yields a simultaneous change in the polarity of all in-gap states (from hole-like to electron-like), as well as the possibility of NDC that is discussed in Supplementary Note 2.

We also note that there exists an additional variation of the MCQPT which in principle yields a discontinuous in-gap LDOS. Namely, the system can undergo two consecutive QPTs of the ground state, from $|0,+,+\rangle$ into $|1/2,-,+\rangle$, and subsequently, from $|1/2,-,+\rangle$ into $|1,-,-\rangle$. Far away from the transition, the level ordering is analogous to that of the scenario discussed so far, and crucially, the impurity also experiences a large change of the occupation (see Fig.~S\ref{fig:supp_spec} (d-f)). There exists again a wide region in the parameter space in which these two QPTs are sufficiently close in $\overline{\varepsilon}$, so that they result in an in-gap LDOS indistinguishable from the scenario discussed above, i.e., with a crossing observed within the energy resolution.

\begin{figure}
	\centering
	\includegraphics[width=\textwidth]{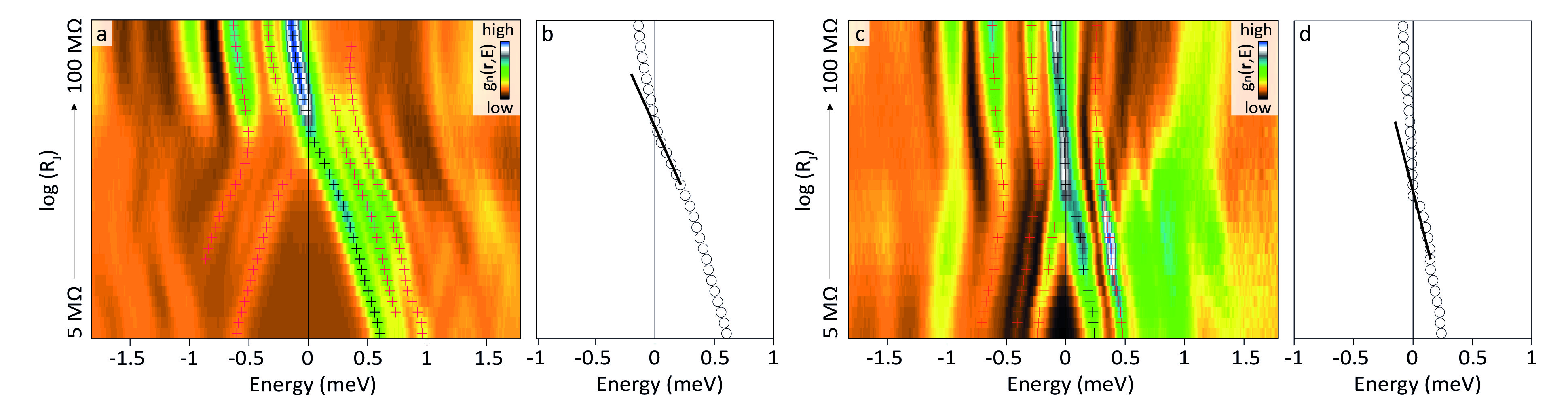}
	\caption{\label{fig:Sslopes} \textbf{Slope change upon crossing $E$ = 0}. \textbf{a} Normalized differential conductance spectra taken at different junction resistances on the same excess Fe atom as in main text Fig. 2. Crosses are fitted locations of the most prominent peaks. The junction resistances run from 5-100 M$\Omega$ on a logarithmic scale. \textbf{b} Fitted peak voltages for the in-gap state that crosses $E$ = 0. The line indicates the slope of the in-gap energy versus junction resistance after crossing $E$ = 0: upon crossing, the slope changes. \textbf{c}, \textbf{d} Same as a, b for another excess Fe impurity. Again, upon crossing $E$ = 0, the slope changes.}
\end{figure}

Finally, we reiterate that since the single-particle excitations in the in-gap LDOS result from different eigenstates before and after the MCQPT, in general there is a discontinuity of slope of the in-gap state energy at the point it crosses zero-bias. As Fig.~S\ref{fig:Sslopes} shows for two different Fe impurities, this is indeed the case in our experimentally observed MCQPT.

\section*{Supplementary Note 2}
In this Supplementary Note we present a minimal transport calculation to account for the negative differential conductance (NDC) observed in the STM measurements. As it was noted in Ref. [\citenum{thielmann_super-poissonian_2005}], NDC in multi-channel quantum-dot like systems originates from a combination of the Coulomb interaction and an asymmetrically-suppressed coupling to the reservoir. We claim that some of the in-gap bound states are comparably weakly coupled to the STM tip, therefore, their spectral footprint in the multi-channel Anderson Impurity model (MCAIM) becomes an NDC peak in the transport calculation.

\begin{figure}
	\centering
	\includegraphics[width=\textwidth]{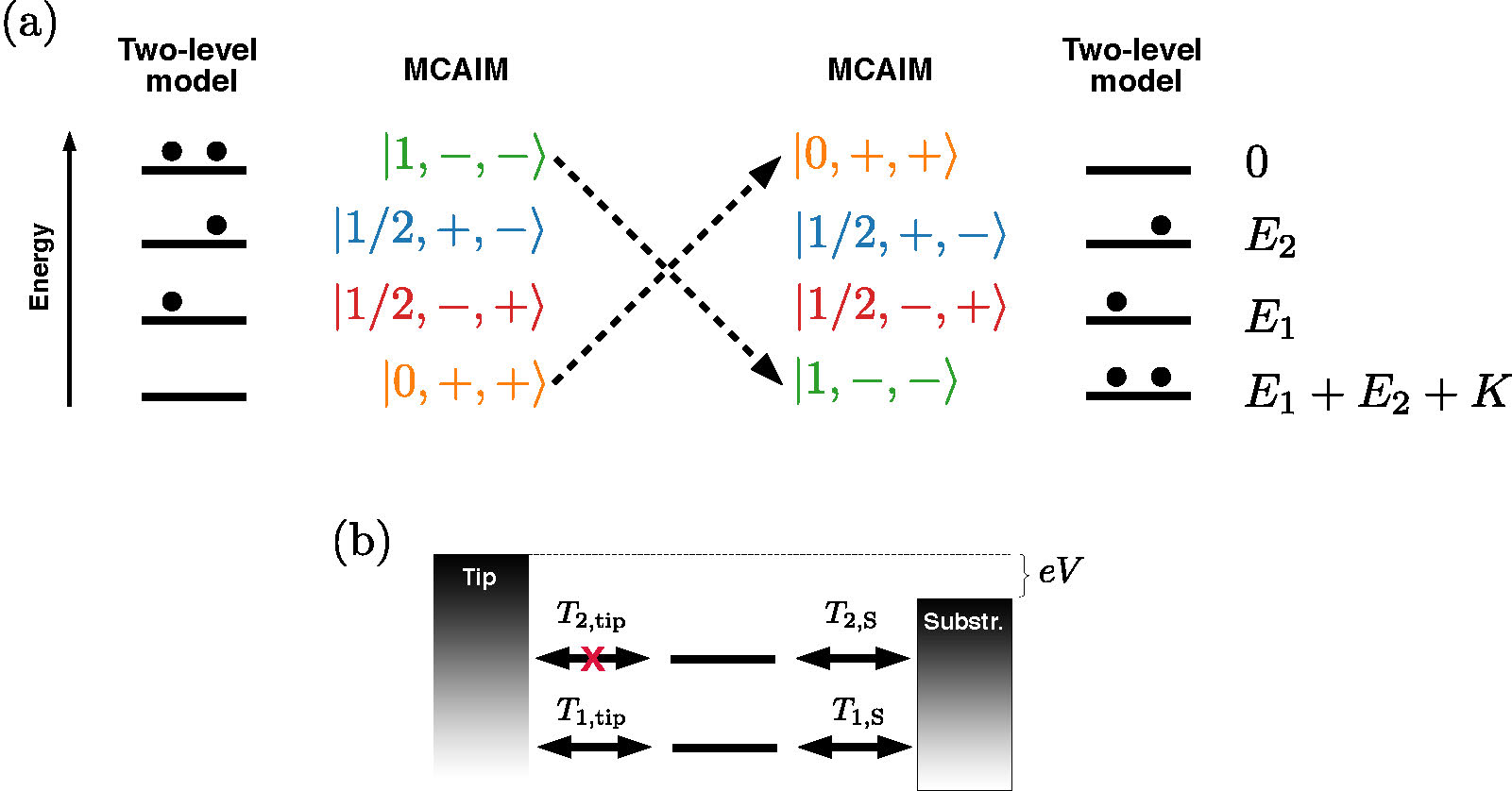}
	\caption{\textbf{Effective two-level model to account for the NDC}. \textbf{a} Schematic mapping between the energy levels in the effective model and the lowest-lying many-body eigenstates in the multi-channel Anderson impurity model. The arrow crossing indicates the MCQPT. \textbf{b} Schematic of the transport model. The coupling between the left reservoir and level 2 is strongly suppressed.}
	\label{fig:supp_ndc_cartoon}
\end{figure}

To explain this behavior, we represent tunneling into the in-gap bound states emerging from a complex impurity-superconductor system as tunneling into simple energy levels. Specifically, we map the four lowest-lying many-body eigenstates of the MCAIM presented in the main text to the energy levels of a spinless two-level interacting model,
\begin{equation}
H_{\mathrm{eff}} = \sum_{l = 1,2} E_l \hat{n_l} + K \hat{n}_{1}\hat{n}_{2},
\end{equation}
with $\hat{n}_l$ the particle-number operator. The energy levels in this effective toy model should be interpreted as the in-gap bound states stemming from the magnetic impurity in the superconducting substrate; they should not be confused with the orbital energy levels in the MCAIM employed to explain the very emergence and behaviour of these in-gap bound states. This is a one-to-one mapping, implying that the re-ordering of the many-body states as we tune the system through the multi-channel QPT is reflected in a re-ordering of the energy levels in the effective model (see Fig.~S\ref{fig:supp_ndc_cartoon}). In addition, $K$ should be interpreted as an effective repulsive interaction which penalizes double occupation of the bound states.

\begin{figure}
	\centering
	\includegraphics[width=\textwidth]{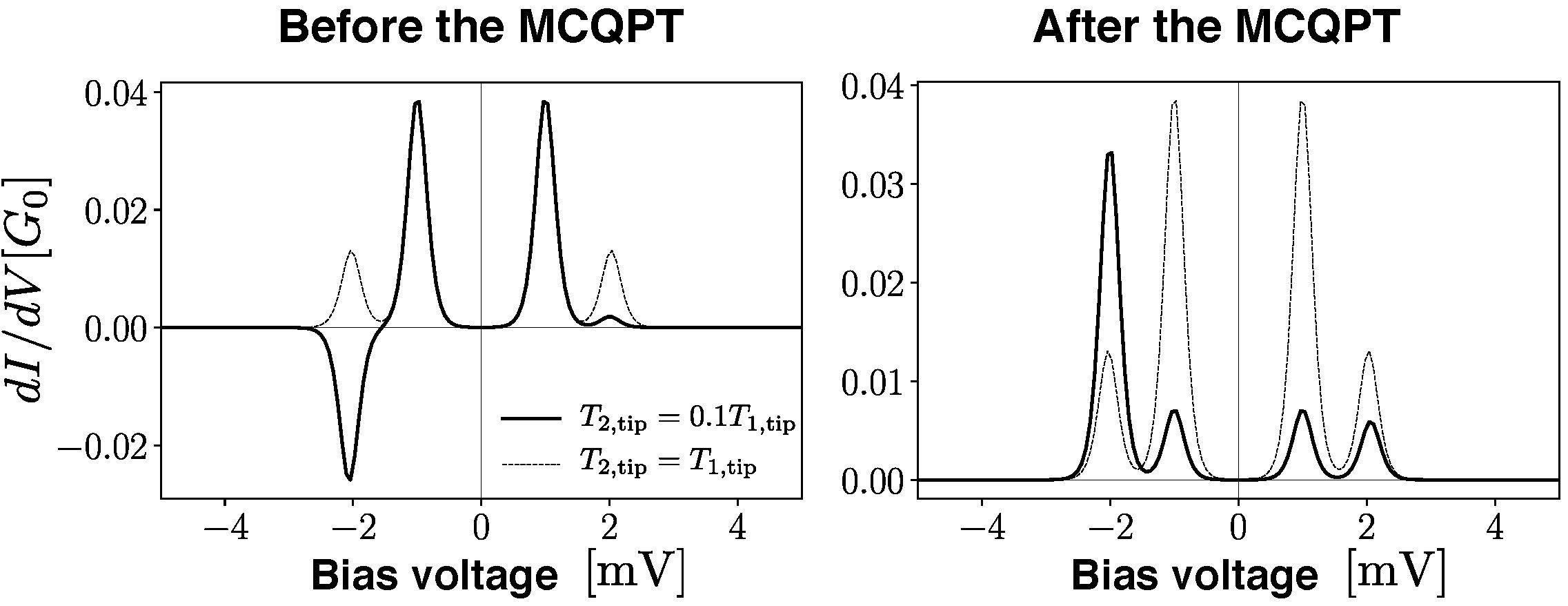}
	\caption{\textbf{Simulated differential conductance in the effective two-level model}. \textbf{a} Level ordering analogous to right-hand side of Fig.~S\ref{fig:supp_ndc_cartoon}. \textbf{b} Level ordering analogous to left-hand side of Fig.~S\ref{fig:supp_ndc_cartoon}. Solid (dashed) lines represent assymetric (equivalent) coupling strengths. Only in the former case there can be NDC. Parameters (in meV): (a) $E_1 = 0.5$, $E_2 = 1.0$, $K=3$, $T_{1,\mathrm{tip}}= 0.005$ (adim.), $k_{\mathrm{B}}T = 0.05$. (b) $E_1 = -4.0$, $E_2 = -3.5$, the rest same as (a).}
	\label{fig:supp_ndc_plots}
\end{figure}

Further, we couple the two-level system to two reservoirs which describe the substrate and the STM tip respectively. Crucially, we assume that one of the couplings between the tip and one of the bound-state wavefunctions is strongly suppressed; this  could be for example due to symmetry reasons. It has been shown experimentally that the bound-state wavefunction shares the same symmetry as the orbital it originates from \cite{choi_ncomm_2017, ruby_orbital_2016}. Therefore, we can easily imagine a small matrix element between the tip wavefunction and one bound-state wavefunction originating from an orbital with an orthogonal symmetry. In the following, we therefore assume ($T_{1,\mathrm{S}} = T_{2,\mathrm{S}} $ and $T_{1,\mathrm{tip}} \gg T_{2,\mathrm{tip}}$). For the sake of simplicity we also assume that $T_{1,\mathrm{S}}=T_{1,\mathrm{tip}}$ but our result do not depend on that latter condition. In Fig.S~\ref{fig:supp_ndc_plots} we present the differential conductance calculated numerically for this effective toy model with a method based on the Lindblad equation, implemented in the python package QmeQ 1.0 \cite{kirsanskas_qmeq_2017}.

Before the transition (Fig.~S\ref{fig:supp_ndc_plots}a), the two lowest-lying peaks in the calculated $dI/dV$ for the effective toy model correspond to single-particle transitions from an empty state to single-occupied levels 1 and 2 respectively. If the tunneling rate from the left reservoir (STM tip) to level 2 (an in-gap bound-state) is much smaller than the rest, electrons cannot flow from level 2. Since the large $K$ penalizes the double occupation of level 1, electrons get stuck in level 2. Therefore, the current collapses and the corresponding excitation appears as a NDC peak at negative bias. After the transition (Fig.~S\ref{fig:supp_ndc_plots}b), the two lowest-lying excitations for the effective toy model correspond to single-particle transitions from a doubly-occupied level to single-occupied levels 1 and 2 respectively. The differential conductance is not negative now, because these transitions involve both levels, therefore, electrons can always escape through level 1.

\begin{figure}
	\centering
	\includegraphics[width=\textwidth]{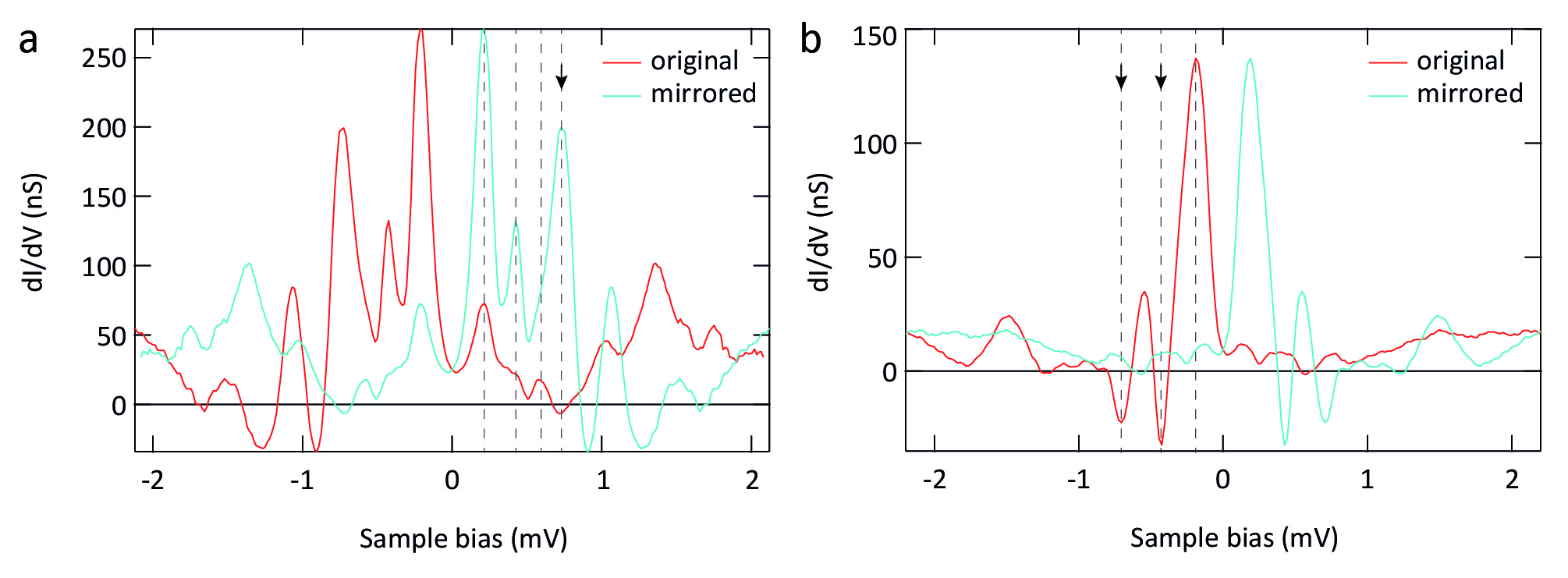}
	\caption{\label{fig:Sspec} \textbf{Negative differential conductance spectra}. \textbf{a}, \textbf{b} Differential conductance spectra taken on different excess Fe impurities. For each spectrum, the differential conductance is plotted as function of the bias voltage (red) and its inverse (blue). The arrows indicate several instances where a NDC dip matches a peak at opposite polarity, in agreement with the two-level model in Fig. \ref{fig:supp_ndc_plots}a. We note that not all NDC dips perfectly match a peak at opposite bias, which likely results from (multiple) partially overlapping states with non-zero width.}
\end{figure}

An important feature of the model above is that the asymmetry in tunneling which causes the NDC does not modify the energies at which the (positive or negative) peaks of the differential conductance occur (see Fig.~S\ref{fig:supp_ndc_plots}). Consequently, the energy $E_{\mathrm{NDC}}$ at which the NDC peak occurs is an excitation energy so that a peak (in principle, either positive or negative) should occur also at $-E_{\mathrm{NDC}}$. Fig.~S\ref{fig:Sspec} demonstrates on two examples of experimental data where indeed such a matching is observed.

The effective toy model captures three key features of the MCAIM. First, excitation transitions before and after the MCQPT are from different ground states into the same excited states in both the effective toy model and the MCAIM. Second, the parity of the effective ground states is the same before and after the transition in both models as well. Third, upon assigning $+1/2$ spin to the particles in the toy model, we also relate the change in the spin quantum number in both models.

Finally, we note that the Lindblad equation method we employed to calculate $dI/dV$ relies on the assumption that all the coupling strengths are much smaller than the temperature, $T_{l,r} \ll k_{\mathrm{B}}T$, with $l=1,2$ and $r=\mathrm{S}, \mathrm{tip}$, which is a reasonable assumption in our experiment. We emphasize again that the couplings $T_{l,r}$ are not directly related to the couplings between the impurity and the superconducting site in the MCAIM $\Gamma_{i,\alpha}$, which are typically larger than $\Delta$ but instead to the intrinsic broadening of the bound states which is typically much smaller than the temperature (see Ref.~[\citenum{perrin_unveiling_2020}] for an experimental estimation of such energy scales).

\section*{Supplementary Note 3}
For completeness, we note that the QPT in the $J_H=0$ case is a transition in a single-orbital AIM model for orbital $a$, from a $s_{\mathrm{tot},a}=0$ ground state with an occupation above one to a $s_{\mathrm{tot},a}=1/2$ ground state with a singly-occupied orbital. Hence, although the spin on the impurity behaves as for two independent Kondo models, the occupation of the impurity changes through the transition, which is not captured in the Kondo model. The departure from the Kondo model is expected \cite{AstAIM} since we are not tuning the coupling to substrate $\Gamma$, but instead the $\overline{\varepsilon}$.